\documentclass{mem}
\usepackage{natbib}
\usepackage{txfonts}
\usepackage{balance}
\usepackage{graphicx}
\usepackage{txfonts}
\usepackage[a4paper]{hyperref}
\idline{79}{3}
\begin{document}
\def\teff{$T\rm_{eff }$}
\def\kms{$\mathrm {km s}^{-1}$}

\title{Extreme Horizontal Branch Stars
}

   \subtitle{}

\author{
U. Heber\inst{1} 
          }

  \offprints{U. Heber}

\institute{
\email{heber@sternwarte.uni-erlangen.de}
}

\authorrunning{U. Heber }

\titlerunning{Extreme Horizontal Branch Stars}

\abstract{A review is presented on the properties, origin and evolutionary 
links of hot subluminous stars which are generally
believed to be extreme Horizontal Branch stars or closely related objects.
They exist both in the disk and halo populations (globular clusters) of the
Galaxy. 
Amongst the field stars a large fraction of sdBs
 are found to reside in close binaries. The companions are predominantly white
 dwarfs, but also low mass main sequence stars are quite common. Systems with
 sufficiently massive white dwarf companions may qualify as Supernova Ia
progenitors. Recently evidence has been found that the masses of 
some unseen companions might exceed the Chandrasekhar mass, hence they must be
neutron stars or black holes.
Even a planet has recently been detected orbiting the pulsating sdB star
V391 Peg.  
 Quite to the opposite, in globular clusters, only very few sdB
binaries are found indicating that the dominant 
sdB formation processes is different in a dense environment. Binary population 
synthesis models identify three formation channels, (i) stable Roche lobe 
overflow, (ii) one or two common envelope ejection phases and (iii) 
the merger of two helium white dwarfs. The latter channel may explain the 
properties of the He-enriched subluminous O stars, the hotter sisters of the 
sdB stars, 
because their binary fraction is lower than that of the sdBs by a factor of ten
or more. The rivalling ''late hot flasher'' scenario is also discussed.
Pulsating subluminous B (sdB) stars play an important role for 
asteroseismology as this technique has already led to mass determinations for a
handful of stars. A unique hyper-velocity sdO star
moving so fast that it is unbound to the Galaxy has probably been ejected by 
the
super-massive black hole in the Galactic centre.             

}
\maketitle{}

\section{Introduction}\label{intro}
Hot subluminous stars are an important population of faint blue stars at high
galactic latitudes closely related to the horizontal branch.
They have recently been studied extensively because they are
common enough to account for the UV excess observed in early-type
galaxies \citep{oconnell99}. Pulsating sdB stars became an important toy for 
asteroseismology
\citep{char04}, and sdB stars in close binaries may qualify as Supernova Ia
progenitors \citep{maxt00,geier07}. 

Subluminous B stars (sdB) have been identified as extreme horizontal branch (EHB)
stars
\citep{heber86}; i.e. they are core helium-burning stars with hydrogen envelopes
that are 
too thin to sustain hydrogen burning (unlike normal HB stars).
Therefore they evolve directly to the white-dwarf cooling sequence by avoiding 
the asymptotic giant branch (AGB).
While the sdB stars spectroscopically form a homogeneous class, a large variety 
of spectra is observed among their hotter sisters, the subluminous O 
stars \citep{heb92,heber08}. 
Most subluminous B stars are helium poor, whereas only a relatively
small
fraction of sdO stars are.

After a brief historical overview, the extreme horizontal branch stars in
globular clusters will be discussed in section \ref{clusters}. Spectroscopic
analyses of the field star population and the role of close binaries are
addressed in sections \ref{field} and \ref{binaries}. Section~\ref{origin} will
focus on the
origin of hot subluminous stars, while section~\ref{pulsations} summarises the 
results from asteroseismology. Section~\ref{kinematics} reports the discovery of
a unique hyper-velocity sdO star. We conclude by briefly presenting 
ideas about the progeny and progenitors of hot subluminous stars.  

\section{Surveys and early results}\label{history} 

The hot subdwarfs were and are being discovered in surveys for UV bright objects
at high Galactic latitudes, starting with the \cite{humason47} survey  
from which the first hot subdwarf stars were identified. Photometric (e.g. 
Palomar-Green) as well as objective prism surveys (e.g. the Hamburg
surveys HS \& HE) have now provided large samples. The Sloan Digital Sky Survey
(SDSS) is a new rich source for hot subdwarf spectra \citep{hirsch08}.
The catalog of hot subluminous stars is available by means of an online 
data base created and maintained by \cite{oestensen06}.

Pioneering spectral analyses of sdB stars in the 1960s revealed strong 
helium-deficiencies, which at first glance seemed to challenge Big Bang
Nucleosynthesis but were soon 
realized to be caused by atmospheric diffusion. The seminal papers by
\cite{newell73} and \cite{green74} set the stage for the field to grow 
driven by new model atmospheres, improved spectroscopy and the advent of UV
spectroscopy. A detailed review on the early development of the field is given
by \cite{lynasgray04}.

\section{The extreme horizontal branch of globular clusters}\label{clusters} 

The literature on extreme or extended horizontal branches in globular clusters
is vast and can not be discussed extensively
here. Instead the reader is referred to the excellent reviews of
\cite{moehler01} and \cite{monibidin08}.

\cite{greenstein71} and \cite{caloi72} were the first to identify the hot HB 
extension observed in very few globular cluster CMDs with the already-known
field sdB stars. 
Various gaps on the blue HBs were found \citep[e.g.
][]{newell76} but their existence 
is now called into question as they may result from poor statistics and 
selection  effects. 
Quantitative spectral analyses in the 1980s \citep[e.g. ][]{heber86b} 
established the similarity 
of GC EHB star to the field hot subdwarfs with respect to all of their 
atmospheric parameters 
(T$_{\rm eff}$,
log~g and He-abundance). The low helium abundance is a consequence of
gravitational settling in the high gravity atmospheres.  


Deep HST photometry increased the number of Globular Clusters showing an EHB
considerably, thereby sharpening the so-called ''Second Parameter Problem''.

The term ''blue tail'' was coined for the most extreme HBs. Far-UV photometry
revealed yet another feature, a very hot ''blue hook'' which seems to extend to below
the helium burning limit \citep{dcruz00}. 

Recent theoretical investigations relate the 
Horizontal Branch morphology to helium enrichment \citep[e.g. ][]{dantona05}.
Previous investigations invoked helium mixing induced by internal 
rotation \citep{swei97}. The recently discovered 
multiple main sequences in $\omega$ Cen have been attributed to helium enriched
sub-populations \citep{bedin04} pointing towards a primordial origin of helium
enrichment. 

The formation of EHB stars in globular clusters may be very different from that
in the Galactic field due to the dense environment in globular clusters.
Dynamical interactions might play an important role for the sdB formation 
through stellar collisions, merging or encounters of binary stars 
\citep{bailyn92}.
Observational evidence is ambiguous. On the one hand, more concentrated or denser
globular clusters have bluer HBs and longer blue tails \citep{buonanno97}
as expected if dynamical interaction is important. On the other hand, however,
no radial gradients have been found even in those clusters which have 
the most strongly populated EHBs 
\citep[e.g.]{bedin00}. In fact, there is no correlation between HB morphology 
and dynamical state of a globular cluster. Pairs of clusters with similar 
HB morphology are dynamically different and, vice versa, others which are
dynamically similar have very different HBs \citep{ferraro97,crocker88}. 
From this point of view, dynamical interaction does not seem to be an important channel for EHB
production in globular clusters (but see section~\ref{binaries}).  

\section{Hot subluminous stars in the field}\label{field}

As many of the field stars are much brighter than their sisters in GCs, they
have been studied in greater detail. The spectral classification ends up in a 
zoo of different subtypes. The O-type class, in particular, is very 
inhomogeneous 
showing vastly different H/He-line spectra, from no helium lines to no Balmer
lines detectable. A similar situation is found for the metal line spectra at
higher spectral resolution. As a consequence no proper classification scheme has
ever been established for the hot subluminous stars.

Appropriate model atmospheres have been developed for quantitative spectral 
analyses allowing to treat on the one hand the metal line blanketing in full 
and
on the other hand deviations from the local thermodynamical equilibrium (in
particular for the O-type stars). 

Quantitative spectral analyses are now available that provide atmospheric
parameters of several hundreds of sdB stars 
\citep[e.g. ][]{saffer94, max01, edelmann03, lisker05}, as well as more than 
100 sdO stars \citep{stroer07,hirsch08}.
. These provide a sound basis to investigate the evolutionary status and origin of
the stars. Metal abundance analyses, however, are still scarce, i.e. available
for a few dozen stars. From UV spectroscopy with HST-STIS a detailed picture of
the abundance pattern of pulsating as well as non-variable sdB stars has been
derived recently \citep{otoole06a}. Abundances for 25 elements 
including the iron group and even heavier elements such as tin and lead have
been derived. Many heavy elements of the
iron group and beyond show large overabundances (by 2--3 dex.) while most
notably iron does not. For the first time the lead
isotopic ratios have be measured to be consistent with the solar ratios
\citep{otoole06b}.     
These results are valuable tools to investigate the diffusion 
processes, i.e. the interplay of gravitational settling, radiative levitation
and a stellar wind. The latter has been found to be an important ingredient 
\citep{unglaub06},
However, it is hard to measure since the expected rates are very low.
Observational evidence is meager yet \citep{heber03b}.

Magnetic fields may be another factor that has been ignored up to now. With the
advent of the ESO VLT, spectropolarimetry has become an option and let to the
first detection of $\approx$kG fields in a few stars \citep{otoole05b}.

\section{Hot subluminous stars in close binaries}\label{binaries}

The fraction of sdB stars in short period binaries (periods less than ten days) is high.
\citet{max01} found 2/3 of their sdB sample were such binaries, whereas 
a somewhat lower fraction of 40\% was found recently for 
  the sample drawn from ESO Supernova Ia Progenitor SurveY
  \citep[SPY, ][]{napi01}. Quite to the opposite, radial velocity 
  variable 
stars are rare amongst the \emph{helium-enriched} sdOs, for which 
\citet{napi04} find that a fraction of radial velocity variables to be
 4\,\% at most.
Obviously, binary evolution plays an important role in the formation of sdB 
stars and possibly also in that of the sdO stars if they are formed by a merging
the components of a close binary system.

This encouraged surveys for radial velocity variable sdB stars in globular 
clusters, which met with little success. The conclusion is that the fraction of
close binary sdB stars is much lower in Globular Clusters than in the field.
\cite{monibidin08} find a binary frequency of only 4\% for the EHB stars in
NGC~6752 indicating that sdB formation in globular cluster may be very different from
that in the field pointing to the relevance of dynamical effects, which might
lead to disruption of binaries or binary mergers. Hence dynamical effects 
should not be disregarded despite of other observational evidence 
to the opposite (see section~\ref{clusters}).
 
The nature of the companions is constrained via the mass function, as well as 
from light variation due to a reflection effect and ellipsoidal deformations.
Up to now the companions have been identified only in the shortest period
systems. Amongst them white dwarfs prevail, but main sequence stars of low mass
are quite common. 

A planetary companion to the pulsating sdB star 
V391 Peg has been discovered 
from sinusoidal variations of its pulsation frequencies \citep{silvotti07}. 
Functionally this is
equivalent to the timing method used to find planets around pulsars.
The discovery of planet around a post-red-giant star demonstrates that planets
can survive the red-giant expansion at distances of less than 2 AU.

At the other end of the mass scale, a few very massive companions have been
found recently. In the most extreme case the companion mass clearly exceeds the 
Chandrasekhar limit and must be a neutron star or a black hole \citep{geier06}. 
The observed fraction of such
companions is much higher than predicted by binary population synthesis models.
Is this a new population of ''hidden'' neutron stars and/or black holes?
  
\section{Evolution and Origin of hot subluminous stars}\label{origin}

The evolutionary status of sdB stars as extreme horizontal branch stars 
appears to be proven beyond
doubt. SdO stars, however, can not be EHB stars, and several options remain for 
their present state of evolution. 

The origin of both sdB and sdO stars remains a puzzle. The main difficulty for 
every scenario is the large amount of mass that has to be lost prior to or at
the start of core helium burning. Canonical evolutionary calculations have been
modified by assuming large mass rates on the RGB leaving us in the
dark about physical mechanism. The occurrence of delayed core helium-flashes
in post-RGB evolution has been discovered as a promising mechanism to produce
sdO stars in particular. The merger of two helium white dwarfs is another 
vital option to explain the
origin of sdO stars rivalling the delayed-core-helium-flash scenario.


Non-standard evolutionary models were introduced to explain the formation of
sdO stars \citep[e.g.][]{swei97,brown01,moe04}. In particular, 
the {\it late hot flasher scenario} predicts that the core helium flash may
occur when the
star has already left the red giant branch (RGB) and is approaching the 
white-dwarf cooling sequence (delayed He core flash). During the flash, He 
and C may be dredged-up to the surface. Hydrogen is mixed into deeper layers 
and  burnt. The
remnant is found to lie close to the helium main sequence, i.e. at the very end
of the theoretical extreme horizontal branch.
The final composition of the envelope is helium-dominated, 
and enriched with carbon  
\citep[or nitrogen if the hydrogen burning during the helium
flash phase burns $^{12}$C into $^{14}$N; ][]{swei97}. 
Indeed, \cite{stroer07} found most of their observed \emph{helium-enriched} 
sdO stars to lie near the model track, suggesting that this scenario may be 
viable. Although it can explain the helium enrichment and 
the line strengths of C and/or N lines as due to dredge up, it fails to
reproduce the distribution of the stars in the T$_{\rm eff}$-$\log g$-diagram 
in detail \citep[see ][]{stroer07}.


The observed high fraction of short period sdB binaries made it clear that 
mass exchange episodes (stable Roche lobe overflow, RLOF, and common envelope 
ejection, CEE)
in close binaries must play an important role for the
origin of sdB stars. 
Recent binary population synthesis study \citep{han03} 
identified three channels 
for forming sdB stars:
(i) one or two phases of common envelope evolution,
(ii) stable Roche-lobe overflow, and
(iii) the merger of two helium-core white-dwarfs.
The latter could explain the population of single stars. Short period binary
white dwarfs
will lose orbital energy through gravitational waves.
With shrinking separation, the less massive object will eventually be 
disrupted and accreted onto its companion, leading to helium ignition.
\citet{saio2000} argue, that this merger product will result in a helium 
burning subdwarf showing an atmosphere enriched in CNO-processed matter.
This scenario therefore can explain these extremely \emph{helium-enriched} sdOs 
showing strong nitrogen lines in their atmospheres.


\cite{castellani93} were the first to point out that helium might not be ignited
in post-RGB evolution depending on the choice of the efficiency parameter
for Reimer's RGB mass loss law. Indeed, there is now observational evidence 
that some sdB stars are not core helium burning objects. In the case of 
HD~188112 the mass has been determined from gravity and an accurate 
Hipparcos parallax to be 0.22 M$_{\odot}$ \citep{heber03a} 
too low to sustain helium burning 
and the star is cooling 
down to the helium white dwarf graveyard. Its post-RGB evolutionary life time is
comparable to that of EHB stars \citep{driebe98}.

First tests of the binary population synthesis models against observations have
been performed from spectra obtained by the SPY consortium for sdB stars 
\citep{lisker05} as well as for sdO stars \citep{stroer07}. 
There are many free parameters to be constrained from 
observations, the most
important being the common envelope ejection efficiency parameter. The
observational tests reject models with very large efficiency parameters 
favoured by
studies of double degenerate stars. In addition there is some evidence against
an uncorrelated mass distribution of the progenitor systems 
\citep[see ][]{lisker05}.
      
\section{Pulsations of hot subluminous stars}\label{pulsations}

Ten years ago, multi-periodic light variations of low amplitudes (a few mmag)
and periods of a few minutes were discovered in sdB stars \citep{kilkenny97}
at almost the
same time they were predicted by theoreticians to be caused by non-radial 
pulsations \citep{charpinet96}. 

Using small
telescopes (typically 1--2\,m) and examining hundreds of sdBs has
yielded more than 30 of these pulsators, each with amplitude
$<$\,50\,mmag \citep[for a review of these objects, now known as V361\,Hya
  stars, see ][]{kilkenny02}. The periods suggest that the stars are $p$-mode
pulsators, but asymptotic theory cannot be applied to the analysis of their
frequency spectra. The possibility of using oscillations to probe the
interiors of sdBs received another boost after the discovery of pulsations
with periods of 45\,min--2\,hr \citep[][ now termed V 1093 Her stars]{green03}. 
The much longer period
pulsations found in these stars indicate they are $g$-modes. The stars
are typically cooler than the $p$-mode pulsators. Theoretical
modelling has found that the pulsations in both groups may be driven by an
opacity bump due to ionisation of iron (and other iron-group elements)
\citep{charpinet01,fontaine03}.
In order for the mechanism to work, iron must be enhanced by diffusion 
processes in the sub-photospheric layers. 
Very sophisticated envelope models have been constructed and have 
met the 
observed distribution of the stars in the V346 Hya stars with great success \citep[see
][ for a review]{fontaine08}.

For the V1093 Her stars, however, models fail to 
reproduce the edges of the instability strip. \cite{jeffery06,jeffery07} 
find that
the results strongly depend on the choice of opacity tables and that the role 
of iron group elements other than iron itself has been underestimated.
     

Of great importance for the development of asteroseismology are the 
so-called hybrid pulsators 
which show both short
period p-mode pulsations as well as long-period g-mode pulsations
\citep[e.g. ][]{schuh06}.
Two puzzling hot subdwarf pulsators have been discovered, one is
a very hot sdO star \citep{woudt06} and the other a sdB stars showing light variations at
periods intermediate between those of the V 361 Hya stars and the V 1093 Her stars
\citep{koen07}. 
While models for
the sdO pulsator are in their infancy, there is as yet no scenario around for 
the latter object.

The main hurdle to overcome towards asteroseismology is the identification of
the observed modes. The period matching technique has already been applied to
half a dozen V 361 Hya stars. Masses and envelope masses have been derived in
this way. In all but one case the results are in good agreement with 
the predicted canonical mass of 0.47 M$\odot$ \citep{randall07}. The period matching technique 
has its limitation as the number of modes predicted to be excited usually is
larger than the number of observed frequencies. 

Other methods have higher predictive power than monochromatic light curves. 
Multi-colour measurements would allow to constrain the mode numbers by its
sensitivity to limb darkening. However, the required accuracy has not yet been
reached for any star to draw conclusion \citep{tremblay06}. 



Stellar surface motions can be derived from radial velocity curves. Pilot
studies \citep[e.g. ][]{otoole05a} have already been successful in detecting  
many modes in velocity that were seen in the light curves including 
combination frequencies. 
Line profile variations can be used
to deduce temperature and gravity variations on the stellar surface.
Despite of the stars' faintness, such analyses have recently been carried out 
for three V361 Hya 
stars \citep{telting04,tillich07}. In all cases, the strongest mode 
has been shown to be a radial one.    
   
\section{Kinematics and population Membership}\label{kinematics}

As the atmospheric abundance patterns of sdB stars are governed by diffusion
processes, they can not be used to establish population membership.
Obviously, some sdB stars belong to the population II because they are found in
globular clusters. For field stars we have to rely on their kinematical 
properties. This becomes now possible for a sufficiently large sample thanks to
accurate radial velocity measurements, spectroscopic distance estimates and
proper motions from several sources. First results indicate that most stars
belong to the thin disk, with a significant fraction of thick disk stars as well
as halo stars \citep{altmann04,richter07}.  

Amongst the sdO stars drawn from the SDSS data base,   
\cite{hirsch05} discovered a so-called hyper-velocity star,
US~708, in the Milky Way halo, with a heliocentric radial velocity of 
+708$\pm$15\,\kms. 

A quantitative NLTE model atmosphere analysis of optical spectra obtained 
with the KECK I telescope
shows that US~708 is a normal \emph{helium-enriched} sdO 
with $T_{\rm eff}$=44\,500~K, $\log(g)=5.25$. Adopting the canonical mass of
half a solar mass from evolution theory the corresponding distance is 19~kpc. 
Its galactic
rest frame velocity is at least 757~\kms, 
much higher than the local Galactic escape velocity (about 430~\kms)
indicating that the star is unbound to the Galaxy.
It has been suggested by \cite{hills88} that such
hyper-velocity
stars can be formed by the tidal disruption of a binary through 
interaction with
the super-massive black hole at the Galactic centre (GC).
Numerical kinematical experiments
are carried out to reconstruct the path of US~708 from the GC. 
US~703 needs about 36~Myrs to travel from the GC to its present
position, which is shorter than its evolutionary lifetime. Hence it is plausible
that the star might have originated from the GC, which can be tested by
measuring accurate proper motions.
A HVS survey has increased the number of known HVS to ten
\citep{brown07}. However, 
US~708 remains the only bona-fide old, low mass HVS star, while all other are
probably young massive stars.     

\section{Conclusions: Progeny and progenitors}\label{progeny}

Subluminous O and B stars evolve directly towards the white dwarf cooling
sequence avoiding the asymptotic giant branch (AGB Manqu\'e). They form an
important channel for low mass white dwarfs. If some sdO stars are formed by
mergers, they are more massive and will evolve into heavier white dwarfs. 
The evolution of close sdB binaries is effected by gravitational wave emission.
SdB stars with M-type dwarf companions probably evolve into cataclysmic
variables. As most of the known systems have periods below the CV period gap, 
this
might be an important channel to form short period CVs. 
SdB binaries with massive white dwarf companions may qualify as progenitors of 
type Ia Supernovae in the context of the double degenerate scenario, if their
combined mass exceeds the Chandrasekhar limit. Two systems are known for which
this is the case \citep{geier07}. 
An evolutionary link to the blue stragglers has been suggested for the sdB stars
in the old open cluster NGC 6791.

Ever since the pioneering work by \citet{green74},
the helium-rich sdO stars were believed to be linked to
the evolution of the hydrogen-rich subluminous B stars. 
Any evolutionary link between subluminous B and O stars, however, is
difficult to explain since the physical processes
driving a transformation of a hydrogen-rich star into a helium-rich one
remain obscure. 

According to the recent studies by \cite{lisker05} and \cite{stroer07} a 
direct evolutionary linkage 
of the hot sdO stars to the somewhat cooler sdB
stars is plausible only 
for the \emph{helium-deficient} sdO stars, i.e. 
the latter are the likely successors to 
sdB stars.

The observed distribution of \emph{helium-enriched} sdO stars is roughly
consistent with the predictions from both the late hot-flasher scenario and 
the helium white-dwarf merger scenario
but do not
match them in detail. 
The occurrence of both a delayed helium core flash and the merger of two
helium white
dwarfs may explain the helium enrichment.
In these cases carbon and/or nitrogen
can be dredged up to the stellar surface, which would
explain the strength of the C and/or N lines in \emph{helium-enriched} sdO 
stars. The lack of close binaries amongst the latter is consistent with a white
dwarf merger
origin.

Self-consistent models of the dredge-up processes during a delayed helium flash
have become available recently \citep{cassisi03}. 
More extensive observational studies are needed to provide accurate metal
abundances, in particular for C and N to test dredge-up models as well as binary
merger models.



%

\bibliographystyle{aa}

\end{document}